\documentclass[12pt]{article}
\usepackage{graphicx}
\usepackage[small]{caption2}
\usepackage{setspace}
\usepackage[super,sort&compress]{natbib}
\usepackage{multirow}
\begin{document}
\begin{center}
{\bf \large Far From Equilibrium Monopole Dynamics in Spin ice}
\vspace{0.5cm}

{C. Paulsen,$^{1 \ast}$ M. J. Jackson,$^{1}$ E. Lhotel,$^{1}$  B. Canals,$^{1}$   D. Prabhakaran,$^{2}$\\
 K. Matsuhira,$^{3}$ S. R. Giblin,$^{4}$ S. T. Bramwell.$^{5}$\\
\normalsize{$^{1}$Institut N\'{e}el, C.N.R.S - Universit\'e Joseph Fourier, BP 166, 38042 Grenoble, France.}\\
\normalsize{$^{2}$Clarendon Laboratory, Physics Department, Oxford University,}\\ 
\normalsize{Oxford, OX1~3PU, United Kingdom.}\\
\normalsize{$^{3}$Kyushu Institute of Technology, Kitakyushu 804-8550, Japan.}\\
\normalsize{$^{4}$School of Physics and Astronomy, Cardiff University, Cardiff, CF24 3AA, United Kingdom.}\\
\normalsize{$^{5}$London Centre for Nanotechnology and Department of Physics and Astronomy,}\\
\normalsize{ University College London, 17-19 Gordon Street, London, WC1H 0AJ, United Kingdom.}\\
\normalsize{$^\ast$ E-mail:  carley.paulsen@grenoble.cnrs.fr}
}

\end{center}

{\bf
Condensed matter in the low temperature limit reveals much exotic physics associated with unusual orders and excitations, examples ranging from helium superfluidity\cite{He} to magnetic monopoles in spin ice\cite{Castel,Ryzhkin}. The far-from-equilibrium physics of such low temperature states may be even more exotic, yet to access it in the laboratory remains a challenge. Here we demonstrate a simple and robust technique, the `magnetothermal avalanche quench', and its use in the controlled creation of nonequilibrium populations of magnetic monopoles in spin ice at millikelvin temperatures.  These populations are found to exhibit spontaneous dynamical effects that typify far-from-equilibrium systems, yet are captured by simple models. Our method thus opens the door to the study of far-from-equilibrium states in spin ice and other exotic magnets.
}

The normal way of controlling the temperature of a system is to connect it thermally to a second body with a much larger thermal mass, which then acts as a thermal reservoir. If it is desired to force thermal excitations out of equilibrium by a rapid temperature quench, then the simplest strategy would be to heat the sample, and then abruptly remove the heating so that the sample is cooled rapidly by the reservoir.  However direct heating of the sample will also tend to heat the reservoir, and this becomes a particular problem in low temperature devices: for example, in a $^3$He-$^4$He dilution refrigerator it may entail heating of the mixing chamber, and ultimately limit the speed of any thermal quench that can be practically achieved. Our technique gets round this problem by using the natural tendency of magnets to undergo magnetothermal `avalanches'  at low temperature \cite{Paulsen95,Paulsen95a, Subedi,Slobinsky10}. It is illustrated in Fig. 1 and discussed further in the supplementary information (1.4). The essential principle is that magnetic work done on the sample is abruptly converted into internal heat, that causes a sudden increase in temperature \textit{inside the sample}. The sample then finds itself at relatively high temperature but connected to a cold thermal bath. The ensuing quench is as efficient and rapid as possible as it  involves minimal heating of the sample environment, which remains at the reservoir temperature, $T$.  

\begin{figure}
\includegraphics[keepaspectratio=true, width=17cm]{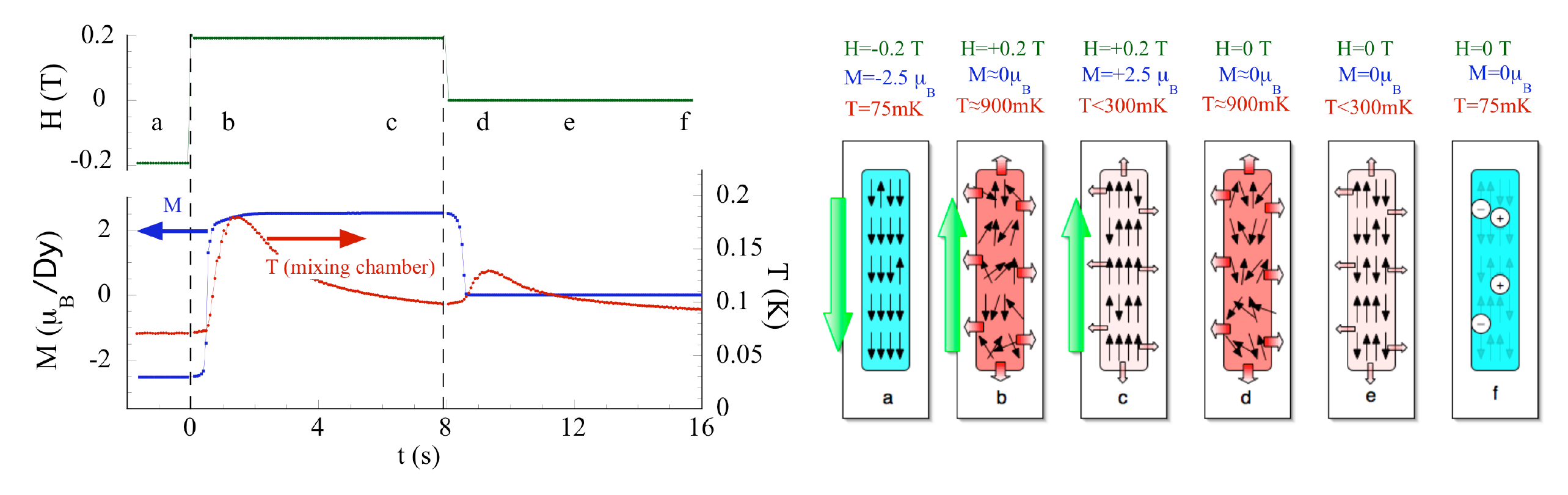}
\caption{{\bf Schematic of the Avalanche Quench (AQ) technique.} ({\bf left}) The applied field $H$, magnetization $M$ and temperature $T$ (of the mixing chamber) as a function of time during the avalanche quench protocol. ({\bf right}) Diagram of the situation inside the sample at various time.
(a) before the procedure begins, the sample was first magnetized to -2.5 $\mu_B$/Dy in a field of -0.2 T, and the temperature of the mixing chamber and sample is stabilized to 75 mK. (b) at time $t=0$, the field is rapidly reversed from -0.2 to +0.2 T at a rate of 0.55 T/s. The magnetization follows, albeit with some delay. The magnetic Zeeman energy $\Delta M\cdot H$ released from the spins rapidly heats the interior of the sample to approximately 900 mK, and the heat then leaks out of the sample to the mixing chamber, as seen as a spike in the temperature on our thermometer.
(c) the mixing chamber remains cold, so the sample cools extremely quickly, limited only by thermal conduction to the Cu sample holder. We estimate cooling to be as fast as 0.07 K/s at 500 mK.
(d) At $t=8$ s, the field is then removed, and the sample avalanches again with less energy (against its internal field). Nevertheless, the temperature inside the sample again approaches 900 mK. (e) the heat is swiftly evacuated to the (still cold) mixing chamber and $M$ approaches 0.
(f)  the result of the fast magnetothermal quench is to freeze in a very large, non-equilibrium density of defects, or monopoles. 
} 
\label{fig_1}
\end{figure}
 
Magnetothermal avalanches typically occur at low temperature ($T<1$ K), a regime also notable for the occurrence of exotic magnetic states based on long range interactions, quantum effects and magnetic frustration~\cite{Harris,Lee,Balents, Lees,Rosenbaum,Tom,Nakastuji}. Hence the avalanche quench technique could be generally used to drive such systems out of equilibrium. We focus on the case of spin ice, a nearly ideal realisation of a magnetic ice-type or vertex model~\cite{Harris}. The far from-equilibrium physics of vertex models is a subject of great intrinsic interest~\cite{Letitia}, so to access it in an experimental system would open up numerous opportunities for the study of new physics. 

 In spin ice materials such as Dy$_2$Ti$_2$O$_7$ and Ho$_2$Ti$_2$O$_7$, the frustrated pyrochlore lattice geometry and local crystal field combines with a self-screening dipole-dipole interaction to give a local `ice rule' that controls low energy spin configurations~\cite{Harris,BramGing,ging}. Simply stated, the minimum energy state corresponds to two spins pointing in and two spins pointing out of each tetrahedron, which stabilises a degenerate and disordered magnetic ground state. 
Thermally generated defects (i.e three spins in and one out, or three out and one in) in the disordered ice rules manifold take the form of effective magnetic monopoles~\cite{Castel, Ryzhkin, Jaubert}. At low temperatures these defects are thermally activated with density in zero applied field varying as $2 \exp(\mu/kT)$ where $\mu$ is the monopole chemical potential ($\mu/k \approx -4.5$ K for Dy$_2$Ti$_2$O$_7$). The monopole description replaces the conventional description based on the spin degrees of freedom, with a model of a symmetric magnetic Coulomb gas with magnetic charge ($\pm Q$). However the Coulomb gas is unusual in that the magnetic monopoles are connected by a network of `Dirac strings'~\cite{Jaubert}.

Below about 0.6 K certain degrees of freedom in spin ice fall out of equilibrium on experimental time scales~\cite{SnyderSchiffer,Matsuhira,Yaraskavitch}. This has been discussed in terms of intrinsic kinetic effects~\cite{BramwellPhilTrans}, slowing down of the monopole hop rate~\cite{Jaubert,Revell,Bovo}, and the trapping of monopoles on extrinsic defects~\cite{Revell}, and it is likely that all such factors play a role. However the degree of non-equilibration has not hitherto been brought under experimental control. Castelnovo {\it et al }~\cite{quench} theoretically showed that a fast thermal quench in the `dipolar spin ice' (DSI) model could create monopole-rich states at low temperature. They also identified the important effect of `noncontractible' monopole-antimonopole pairs, for which annihilation is itself a thermally activated process that would lead to a dynamical arrest at low temperature.

\begin{figure}[h]
\includegraphics[keepaspectratio=true, width=16cm]{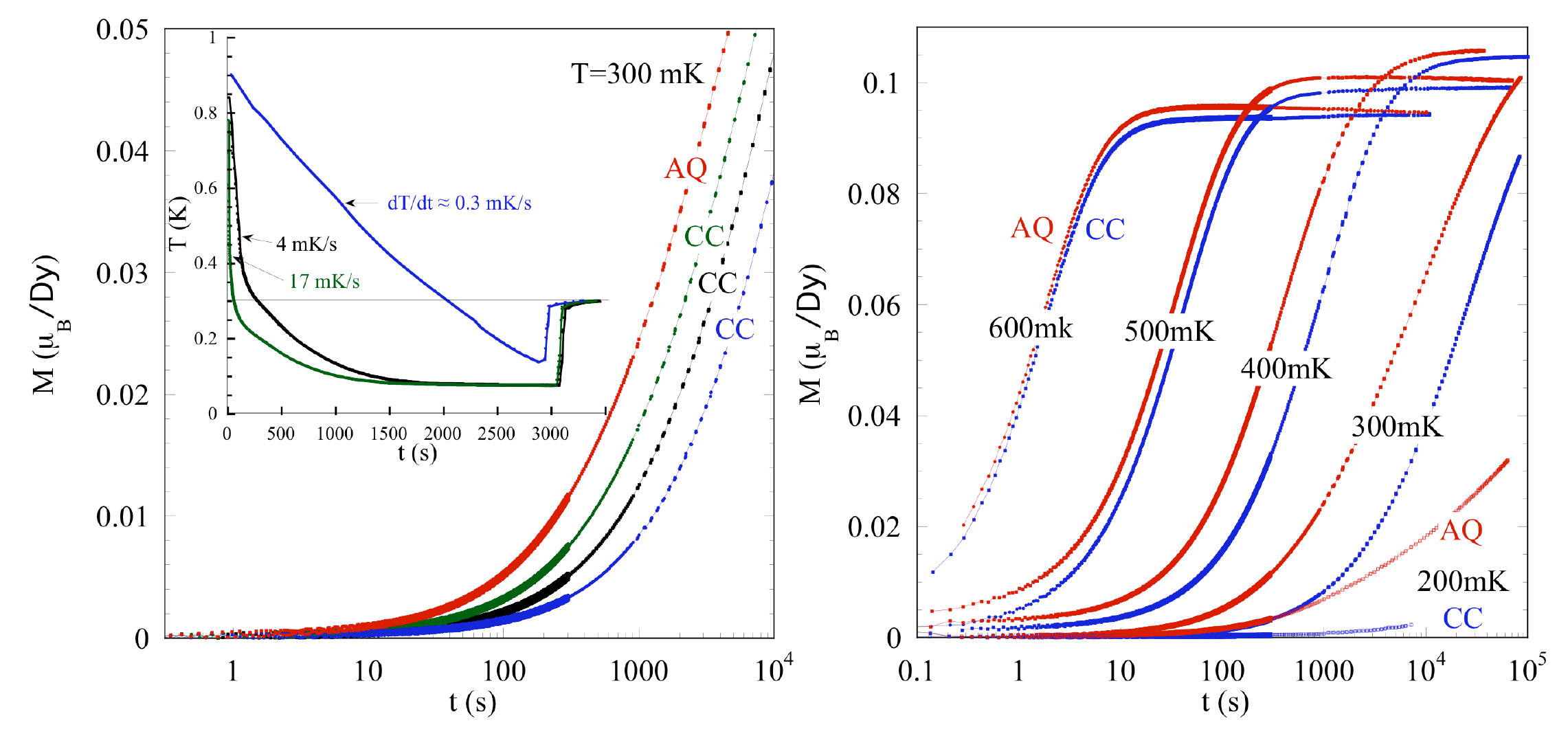}
\caption{{\bf Relaxation of the magnetization $M$ vs time $t$.} Measured at 300 mK  ({\bf left}), three of the relaxation runs were first prepared using the conventional cooling (CC) routine with three different cooling rates as shown in the inset, (blue: $r\approx0.3$ mK/s, black: $r\approx4$ mK/s and green: $r\approx17$ mK/s) and one relaxation curve (red: $r \approx 70$ mK/s) using the avalanche quench protocol outlined in Fig. 1. The applied field was 5 mT for all of the curves. ({\bf right}) Pairs of CC (blue: $r\approx0.3$ mK/s) and AQ (red) relaxation curves measured at fixed temperatures ranging from 600 to 200 mK. The faster the cooling rate, the faster the relaxation, and clearly the AQ protocol is the fastest. The trend becomes very marked below 300 mK. } 
\label{fig_2}
\end{figure}

The clear importance of the rate at which the sample is cooled ($r=dT/dt$) on the subsequent relaxation at low temperature is illustrated in Fig. \ref{fig_2}. Three of the curves shown in this figure followed the conventional cooling protocol to prepare the sample: (1) the sample was first heated to 900 mK for approximately 1 min in zero magnetic field (and the measured magnetization was $M=0$) (2) the sample was then cooled at 3 different rates as shown in the inset. (3) After a total cooling/wait period of 3000 s, the temperature was regulated to 300 mK for another 600 s. (4) Then a 5 mT field was applied and the clock was reset and the ensuing relaxation of the magnetization recorded. It is seen in Fig. \ref{fig_2} that the faster the cooling rate, the faster the magnetic response.  Also shown in the figure is relaxation data taken after the sample was prepared using the magneto-thermal avalanche quench (AQ) protocol outlined in Fig. \ref{fig_1}. According to the model of Ryzhkin,  ${\bf J} = \partial {\bf M}(t)/ \partial t$ is the monopole current density and as monopole dynamics are in the diffusive limit, a larger current density directly implies a larger monopole density~\cite{Ryzhkin}. Thus we conclude that both conventional and avalanche cooling protocols  result in the creation of non-equilibrium monopole populations. However Fig. 2 shows how the avalanche quench leads to significantly faster relaxation than the fastest conventional protocol, and hence we may infer that it stabilizes a proportionately larger monopole density. Furthermore, the difference between the two protocols becomes more obvious at lower temperature as seen in right panel. Note that, although there are differences in the relaxation curves, both cooling protocols lead to the same final magnetization after application of a field, suggesting that the final state is an equilibrium state.   The data presented here were performed with the field applied along the [111] direction. Our results are qualitatively similar when the applied field was  perpendicular to the [111] axis and along the [110] axis (see ``supplementary information, 1.3''). 

\begin{figure}[h!]
\includegraphics[keepaspectratio=true, width=16cm]{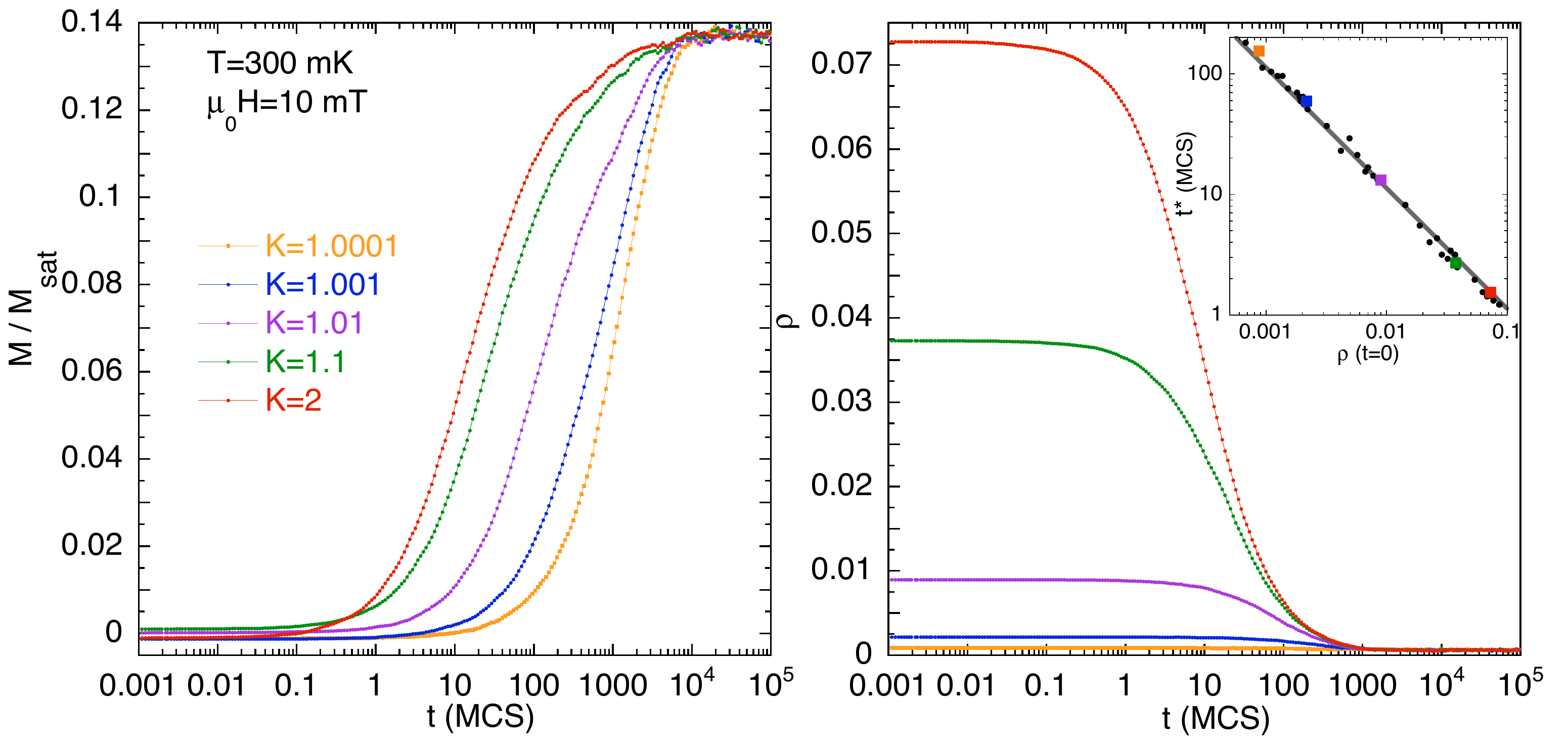}
\caption{{\bf Simulation of non-equilibrium effect.} Magnetization $M$ ({\bf left}) and monopole density $\rho$ ({\bf right}) vs time $t$, from a kinetic Monte Carlo simulation of near neighbour spin ice - a 16 vertex model, similar to that studied in Refs. \citenum{quench, Letitia}. The inset shows the time $t^*$ needed to reach the magnetization $M(t^*)=M(T,t=\infty)/10$ as a function of the density of monopoles $\rho$ at $t=0$, which depends itself on the simulated cooling rate ($K$). The line ($\propto 1/\rho(t=0)$) is a guide to the eye.
The numerical system assumes single spin flip dynamics with periodic boundary conditions on a pyrochlore lattice of size $L\times L \times L \times16$ ($L=8$ in this figure). The model is thermally quenched by reducing the temperature in equal logarithmic steps, from 1 K to 300 mK, in zero applied magnetic field. The logarithmic decrement, from right to left is $K=$1.0001, 1.001, 1.01, 1.1 and 2. 
A field of 10 mT is then applied at $t=0$ and $M$(t) is monitored until it reaches its asymptotic, cooling rate independent, value. The microscopic parameters for the simulation are those of Dy$_2$Ti$_2$O$_7$ (from Ref \citenum{Jaubert}). Each $M$(t) curve corresponds to an average of 100 independent cooling scenarios. Time is given in units of Monte Carlo Steps (MCS), 1 MCS being associated to a stochastic sampling of the whole sample.
(Note that similar simulations with the DSI model reproduces the same behaviour. See ``supplementary information - 1.5''.)} 
\label{fig_3}
\end{figure}

Remarkably, we found that the dependence of the relaxation on the cooling rate is captured by the most simple model of spin ice: the near neighbour spin ice model (simulations of the dipolar model \cite{ging} are reported in the ``supplementary information, 1.6''). Fig. \ref{fig_3} shows the results of a kinetic Monte Carlo simulation of near neighbour spin ice - a 16 vertex model, similar to that studied in Refs. \citenum{quench, Letitia}. The numerical system assumes single spin flip dynamics with periodic boundary conditions and was thermally quenched by reducing the temperature in equal logarithmic steps. 
It is clear that it qualitatively reproduces experiment. In addition, the simulation correlates the rate of the relaxation of $M(t)$ with the $t=0$ monopole density (see inset of Fig. \ref{fig_3}), which is itself related to the $dT/dt$ cooling rate of the sample: the faster the cooling rate, the more monopoles, and the more monopoles the faster the initial relaxation. 
As anticipated, the relaxation rate is roughly inversely proportional to the initial monopole density (see ``supplementary information, 1.3-1.4'').

The qualitative success of the near neighbour spin ice model also shows that the avalanche quench has realized intrinsic  non equilibrium behaviour in spin ice. While extrinsic defects can influence the dynamics of spin ice at low temperature~\cite{Revell}, it is clear that they do not cause the behaviour we observe (see ``supplementary information, 1.3''). The irrelevance of extrinsic defects to our experiments are consistent with creation of monopole rich states, in which the monopole density far exceeds that of extrinsic defects.

Close inspection of the experimental curves (see right panel of Fig. \ref{fig_2}), reveals an intriguing behaviour of the AQ magnetization at long times. All the conventional cooling protocols lead to a monotonic, roughly exponential  evolution of $M(t)$, typical of systems whose initial state is not too far from equilibrium.  In contrast, the avalanche quench leads to a non-monotonic $M(t)$ which initially increases, but later reaches a maximum before finally {\it decreasing} towards the final equilibrium state. In thermodynamics the magnetization is an increasing function of field, so the behaviour following avalanche quenching suggests that the magnetic system is exploring states that are sufficiently far from equilibrium for oscillatory behaviour to occur.  It therefore seems that the avalanche quench puts the monopole gas far-from-equilibrium in this sense, while the conventional cooling leads to near-equilibrium behaviour. Interestingly, a similar overshoot appears in the simulations for small system sizes when considering open boundary conditions, but it disappears in the thermodynamic limit.

A plausible explanation for the enhanced relaxation and overshoot of the magnetization has its basis in the observation that the AQ protocol freezes in a much larger, non equilibrium density of monopoles, hitherto inaccessible at very low temperatures. When the magnetic field is applied, the AQ prepared sample will have a tremendous advantage over the slow conventionally cooled sample, at least in the beginning, because it does not need to create new monopoles (a process that becomes exceedingly slow at low temperatures due to the high energy barrier for their creation). Thus the existing monopoles are free to hop between tetrahedra within the constraints of the ice-rules  leaving behind Dirac strings of overturned spins. The magnetization grows, and the process is so efficient, that the magnetization can even overshoot the thermal equilibrium value.  However at longer times, the lower entropy of the strings and finite temperature will favor breaking of the strings and an eventual rearrangement of the spins toward true thermodynamic equilibrium. 

\begin{figure}[h!]
\includegraphics[keepaspectratio=true, width=16cm]{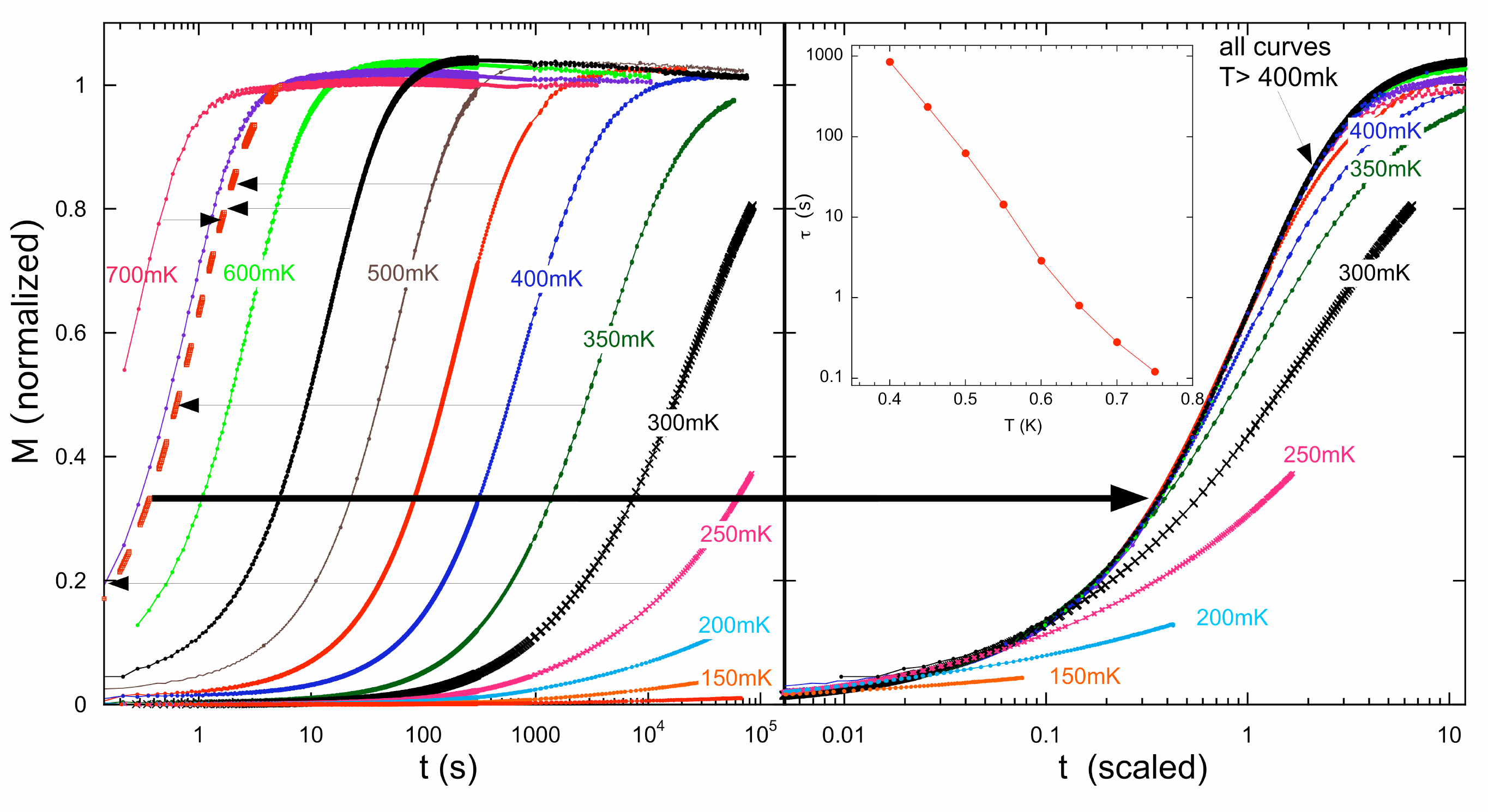}
\caption{{\bf Scaling of the relaxation.} The normalized AQ relaxation $M(T,t)/M(T,t=\infty)$ measured from 700 to 75~mK in a 5 mT field ({\bf left}). The dashed line is the 450 mK curve with the time scaled (i.e. shifted to the left) in such a way that $M= 1-1/e$ occurs at 1 s. The arrows indicate similar scaling for other curves. The scale factor represents an average relaxation time $\tau$ and is shown in the inset as a function of temperature down to 400 mK. ({\bf right}) The results of the scaling for all of the curves in the left panel. For $T > 0.4$ K all the curves can be more or less superimposed as shown. However at lower temperature, the curves cannot be scaled, indicating that there is a change in the dynamic behaviour, and $\tau$ for these curves cannot be so simply defined. } 
\label{fig_4}
\end{figure}

A final interesting property of the monopole-rich state is illustrated in Fig. \ref{fig_4}. It is shown in the right panel that, above 0.4 K, the relaxation curves $M(t)$ collapse when scaling the time axis, whereas that below 0.4 K they do not. This reveals a basic change of dynamical regime at $T < 0.4$ K, which may indicate either a change in the microscopic spin or monopole dynamics, or else an approach towards a different equilibrium state. The latter is consistent with a recent specific heat study that has suggested the onset of ordering correlations in this temperature range~\cite{Kycia}, and also with neutron scattering evidence that reveals a gradual departure from pure spin ice correlations (although no tendency to order) as Dy$_2$Ti$_2$O$_7$ is cooled to 0.3 K~\cite{Fennell1,Yavorskii}. On the other hand the `dual electrolyte' of the monopole model shows a striking change of dynamics exactly in this temperature range, that is associated with the Wien effect for magnetic monopoles~\cite{steve1,sean,Vojtech}. To distinguish these possibilities will be an object for future work.

 In conclusion, the avalanche quench technique has been used to create a monopole-rich state 
in spin ice. The non-monotonic relaxation of this state coupled with the success of the near neighbour spin ice model in describing the main features of the relaxation, indicates that we have experimentally observed the intrinsic far-from-equilibrium 
dynamics of a 16-vertex model~\cite{Letitia}. The avalanche quench technique that we have described may be potentially applied to other magnetic systems in order to similarly drive them into a far-from-equilibrium regime.

\vspace{1cm}
\noindent
{\bf Acknowledgements} S.R.G would like to acknowledge the support of the European Community - Research Infrastructures under the FP7 Capacities Specific Programme, MICROKELVIN project number 228464.

\newpage

\begin{center} {\bf \large Supplementary Information for \\ Far From Equilibrium Monopole Dynamics in Spin ice} \end{center}
\vspace{0.5cm}

\renewcommand{\thefigure}{S\arabic{figure}}
\renewcommand{\thetable}{S\arabic{table}}

 \setcounter{figure}{0} 

\paragraph*{Materials and Methods.}
\paragraph*{1.1. Samples:}

For the study of the cooling rate dependence on the relaxation, three different crystalline samples of Dy$_2$Ti$_2$O$_7$ were used, two of which were measured along different axes (see table \ref{SampleTable}).
All three were grown by the floating zone method as described in Ref. \citenum{Matsuhira02} (Samples 1 and 3, grown at the Kyushu Institute of Technology) and Ref. \citenum{Prabhakaran11} (Sample 2, grown at Oxford University).

The crystal alignment with respect to the field is accurate to within a few degrees.
The demagnetization factors $N$ (see table~\ref{SampleTable}) were calculated with the analytical form for a rectangular prism \cite{Aharoni98}.

\begin{table}[h]
 \begin{tabular}{|*{5}{c|}}   \hline
 Sample& Dimensions (mm$^3$) & Mass (mg) & Field direction & $N$ (cgs) \\ \hline
 1 & 3.80 $\times$ 1.85 $\times$ 0.90 & 44.2 & [111] (presented in this study) & 1.74 \\
                                                                                                          &&& perpendicular to [111] & 3.65 \\ \hline
 2 & 5.05 $\times$ 1.50 $\times$ 1.50 & 73.4 & [111] & 1.59 \\ \hline
 3 & 3.50 $\times$ 2.13 $\times$ 1.75 & 92.4 & [001] & 2.68 \\  
 &&& arbitrary angle (close to [111]) & 4.77 \\
 \hline
    \end{tabular}
\caption{Sample Details}
\label{SampleTable}
\end{table}

\paragraph*{1.2. Details of the experimental setup:}
All measurements were performed using a low temperature SQUID magnetometer developed at the Institut N\'eel. The magnetometer is equipped with a miniature dilution refrigerator capable of cooling the sample to 70 mK. A unique feature of the setup is that absolute values of the magnetization can be made using the extraction method, without heating the sample. The magnetometer has separate dc and ac coils. The maximum dc field is 4000 Oe. The dc field can be swept at $dH/dt=$ 5500 Oe/s or 8000 Oe/s depending on power supply and feed-back used. There is an associated group of balancing coils that can be adjusted in such a way as to make the field changes nearly undetectable to the squid gradiometer. An active shield screens the magnetic field from an outer superconducting lead shield, and the whole system is inclosed in a mu metal shield leaving only a few milli gauss residual field.  

\begin{figure}
\center{
\includegraphics[keepaspectratio=true, width=8cm]{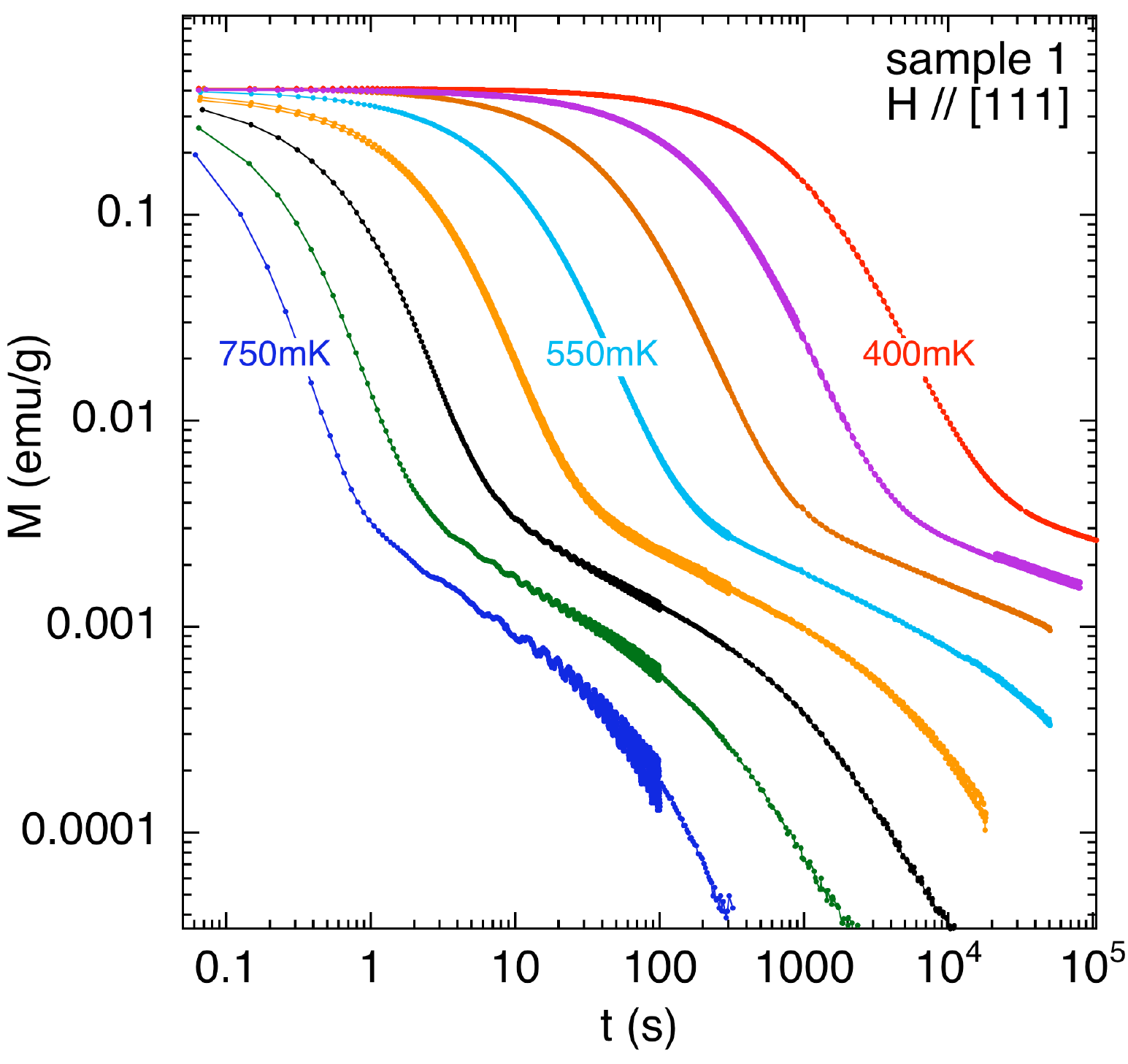}}
\caption{Field cooled Òconventional Ò relaxation of the magnetization for sample 1, $H$ // [111]. These measurements were made by cooling the sample in a field of 10 Oe from 900 mK to base temperature (70 mK) with a cooling rate of approximately $dT/dt=10$ mK/s at 500 mK. After 10 minutes, the temperature was then ramped up to  the respective measuring temperatures shown in the figure, and regulated for another 10 minutes, after which the field was cut, the timer set to zero, and measurements started in the relative mode. At longer times, when the relaxation becomes slower, measurements were then made using the extraction method. For these measurements a small 0.002 Oe field was applied to compensate for the residual earths field after the shielding.} 
\label{fig_S1}
\end{figure}

A measurement by extraction typically takes about 10 s to complete. However, depending on the temperature, the relaxation of the sample can be fast, and after 10 s most of the change in the magnetization could have already happened. Thus a hybrid relative/absolute measurement was adopted.  Relaxation measurements were first made in the relative mode by placing the sample in one of the detection coils. The measuring field was then abruptly applied, and the relative relaxation and the temperature was recorded at approximately 10 points/s for the first 100 to 300 seconds. Then measurements were made by the extraction method, and the initial relative measurements were adjusted by an offset. We estimate errors in the adjustments to be less than 5\% (See figure \ref{fig_S1}).

\begin{figure}
\includegraphics[keepaspectratio=true, height=7.5cm]{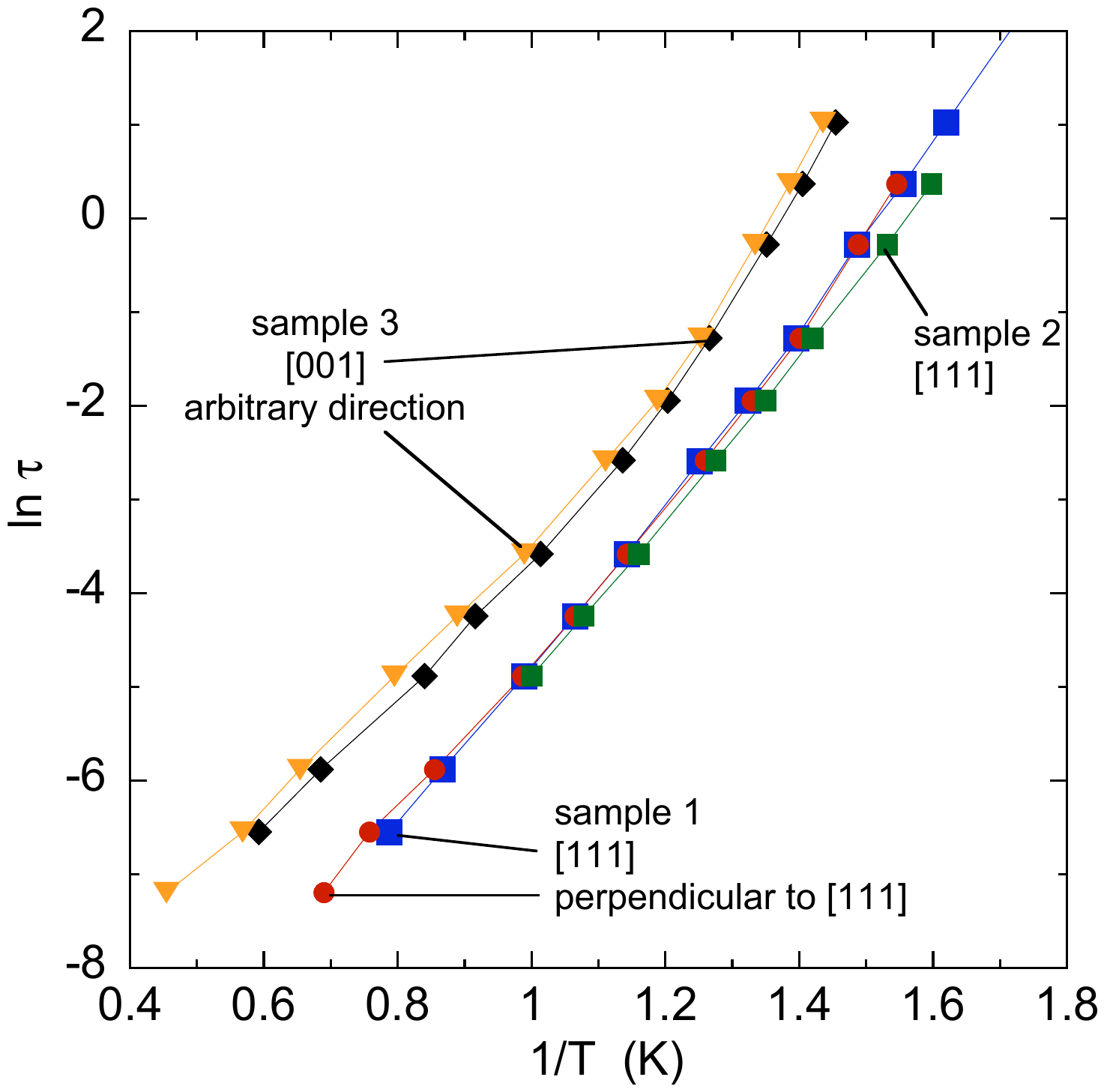}
\includegraphics[keepaspectratio=true, height=7.5cm]{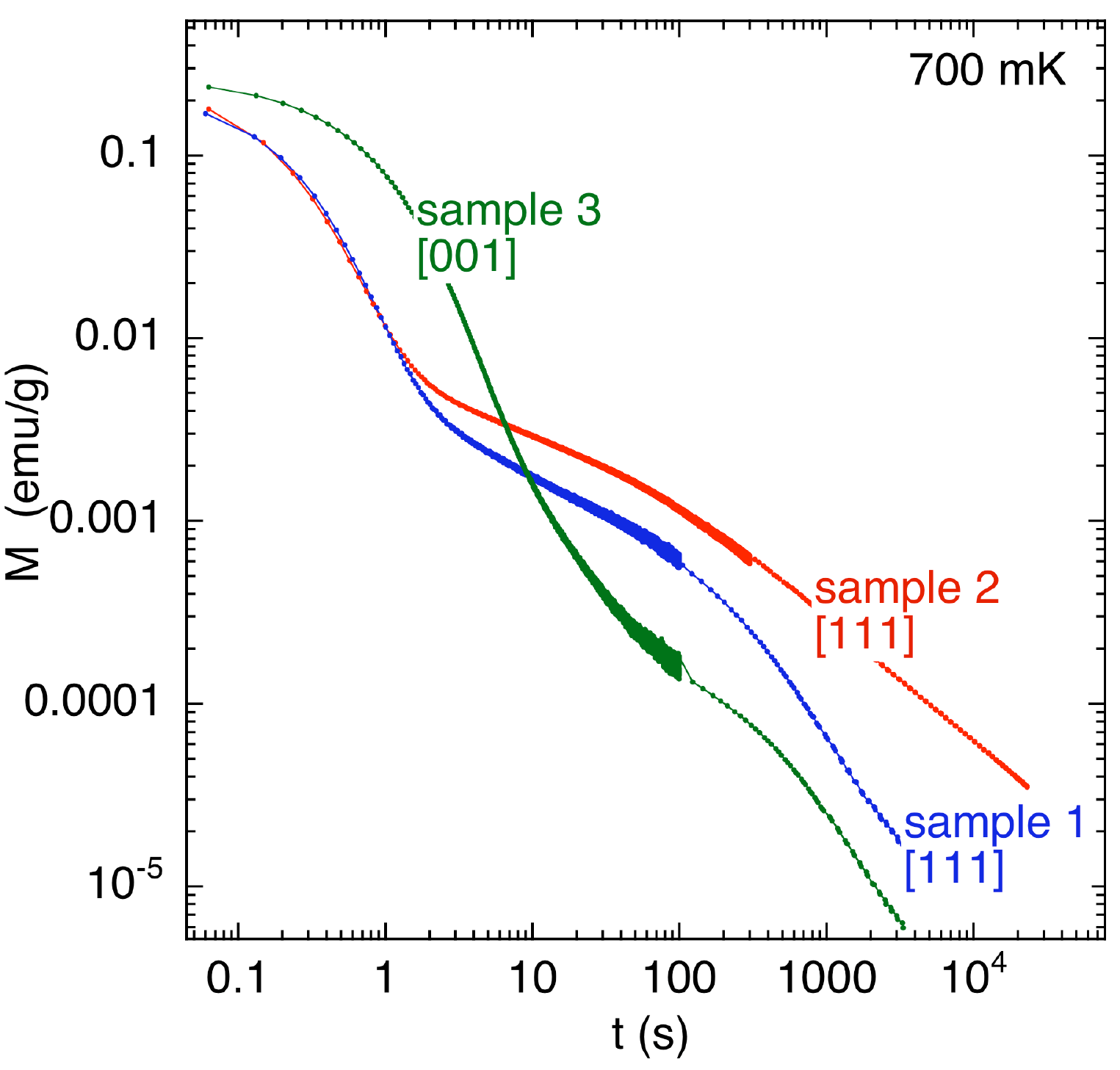}
\caption{({\bf left}) The logarithm of the relaxation time $\tau$ obtained by ac susceptibility plotted against $1/T$ for three different samples, and two different orientations. The $\tau$ were defined using the peak in the (demagnetization corrected) imaginary susceptibility vs temperature. ({\bf right}) Relaxation of the magnetization using the conventional protocol as described in figure \ref{fig_S1} for three different samples, measured at 700 mK.
As can be seen in both plots,  samples 1 and 2 were similar, but sample 3 has a significantly slower relaxation.
} 
\label{fig_S2}
\end{figure}

\paragraph*{1.3. Sample dependence and importance of the cooling rate:}
Although samples 1 and 2 came from different laboratories, their overall relaxation characteristics were very similar. This is in contrast to sample 3, which was significantly slower than samples 1 and 2, as observed by the frequency dependence of the ac susceptibility at high temperature and by relaxation measurements at low temperature, as shown in figure \ref{fig_S2}.

\begin{figure}
\includegraphics[keepaspectratio=true, width=7.5cm]{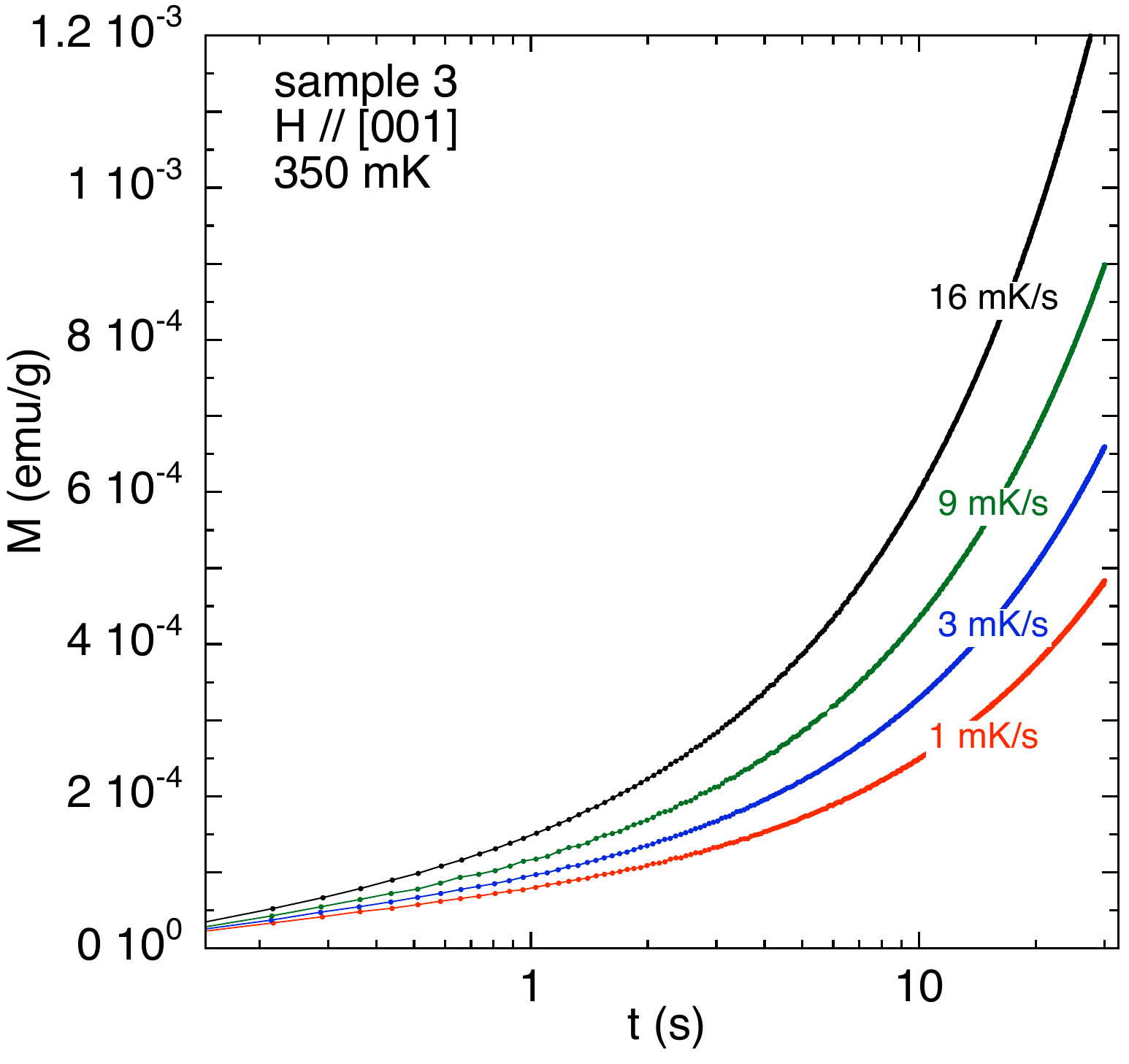}
\includegraphics[keepaspectratio=true, width=7.5cm]{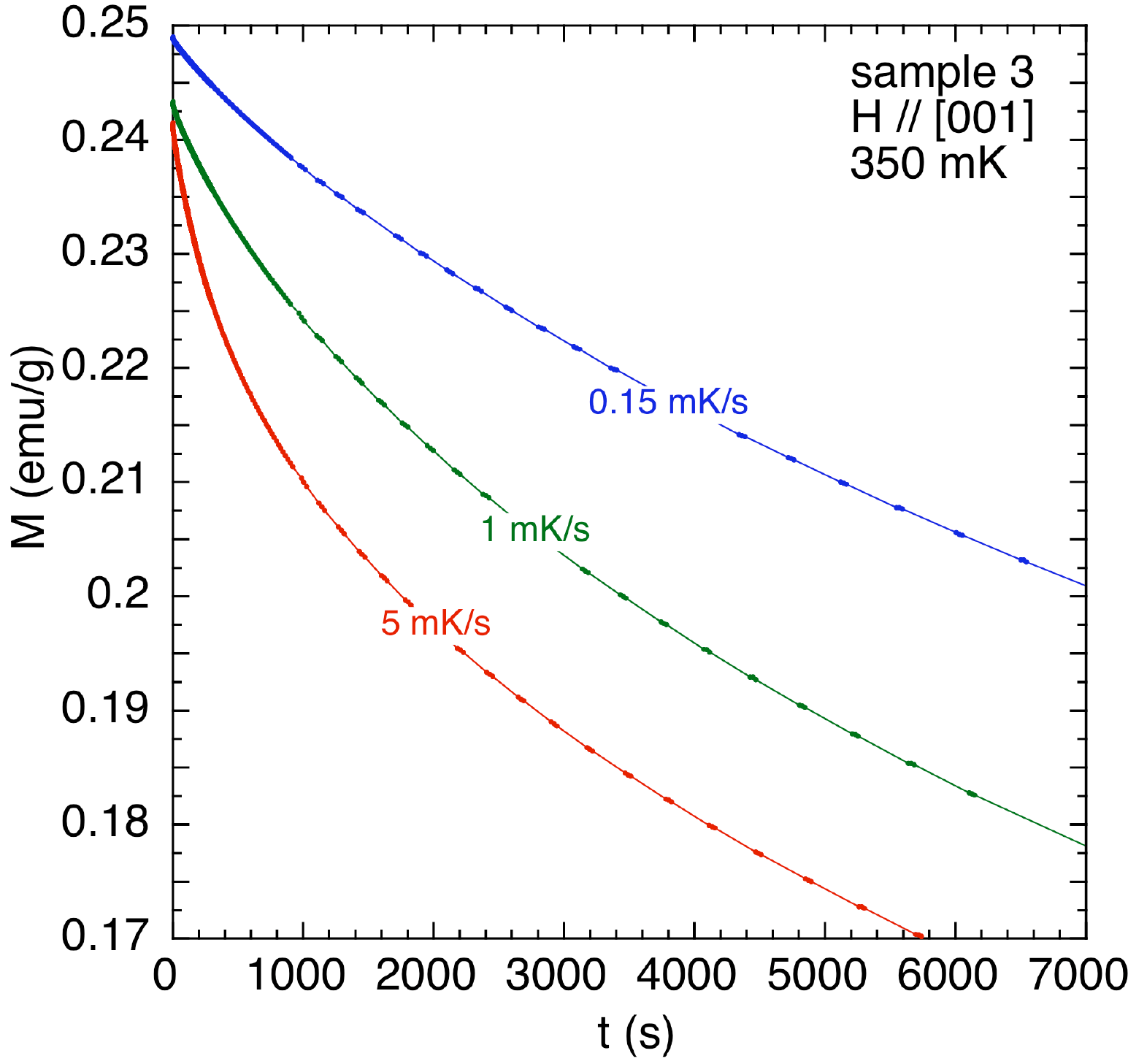}
\caption{Although sample 3, $H$ // [001] was slower than samples 1 and 2, the importance of the cooling rate dependence on the relaxation was the same. Shown in the plots are examples of the relaxation of the magnetization for sample 3 with the field along the [001] direction and measured at 350 mK using the same protocol as described in figure \ref{fig_S1}. ({\bf left}) The sample was first ZFC, and then a 5 Oe field was applied, the cooling rates $dT/dt$ at 500 mK were 16 mK/s to 1 mK/s as indicated in the plot. ({\bf right}) The sample was field cooled in 50 Oe field, with the indicated cooling rates.  } 
\label{fig_S3}
\end{figure}

\begin{figure}[h!]
\includegraphics[keepaspectratio=true, width=8cm]{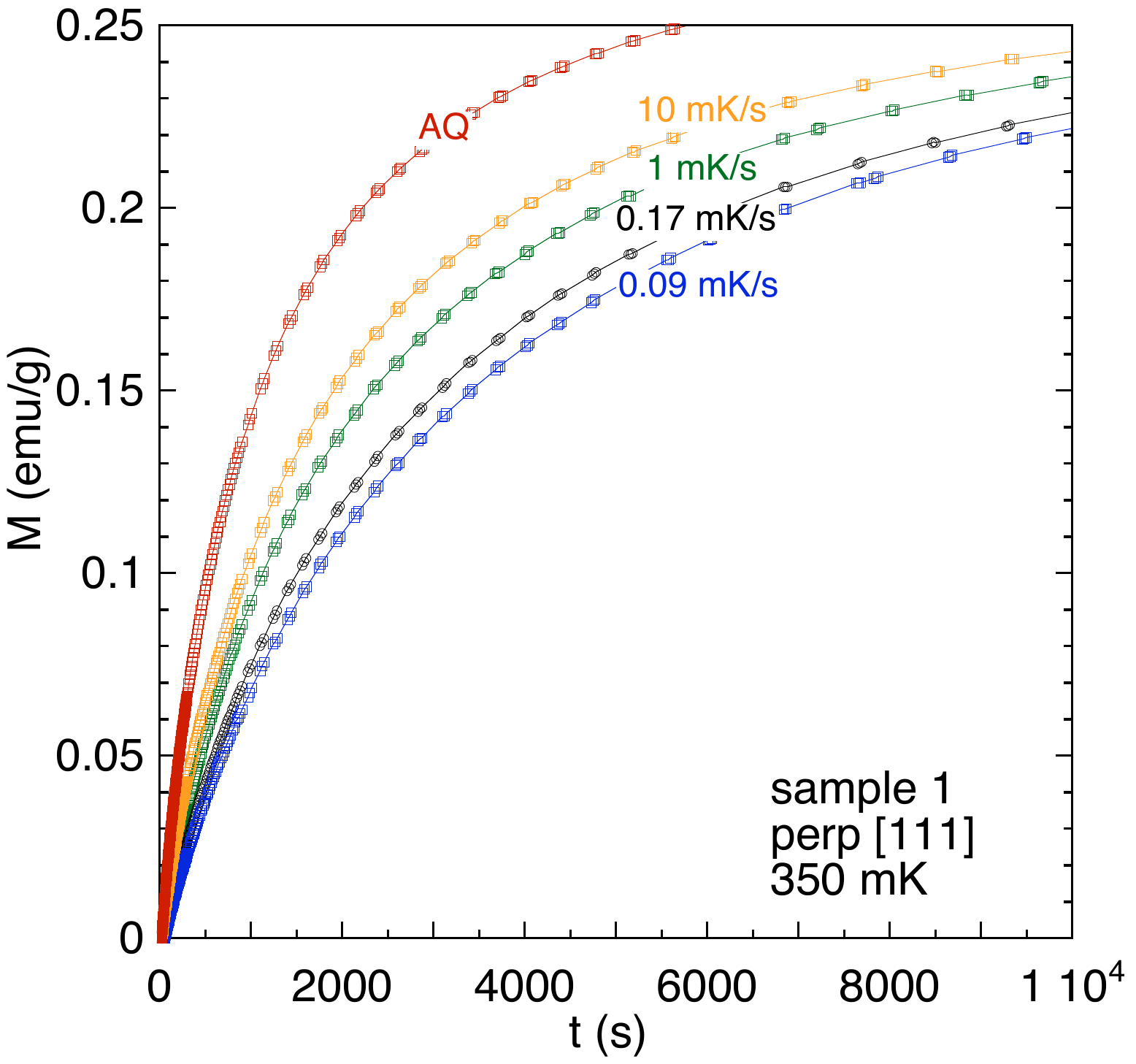}
\includegraphics[keepaspectratio=true, width=8cm]{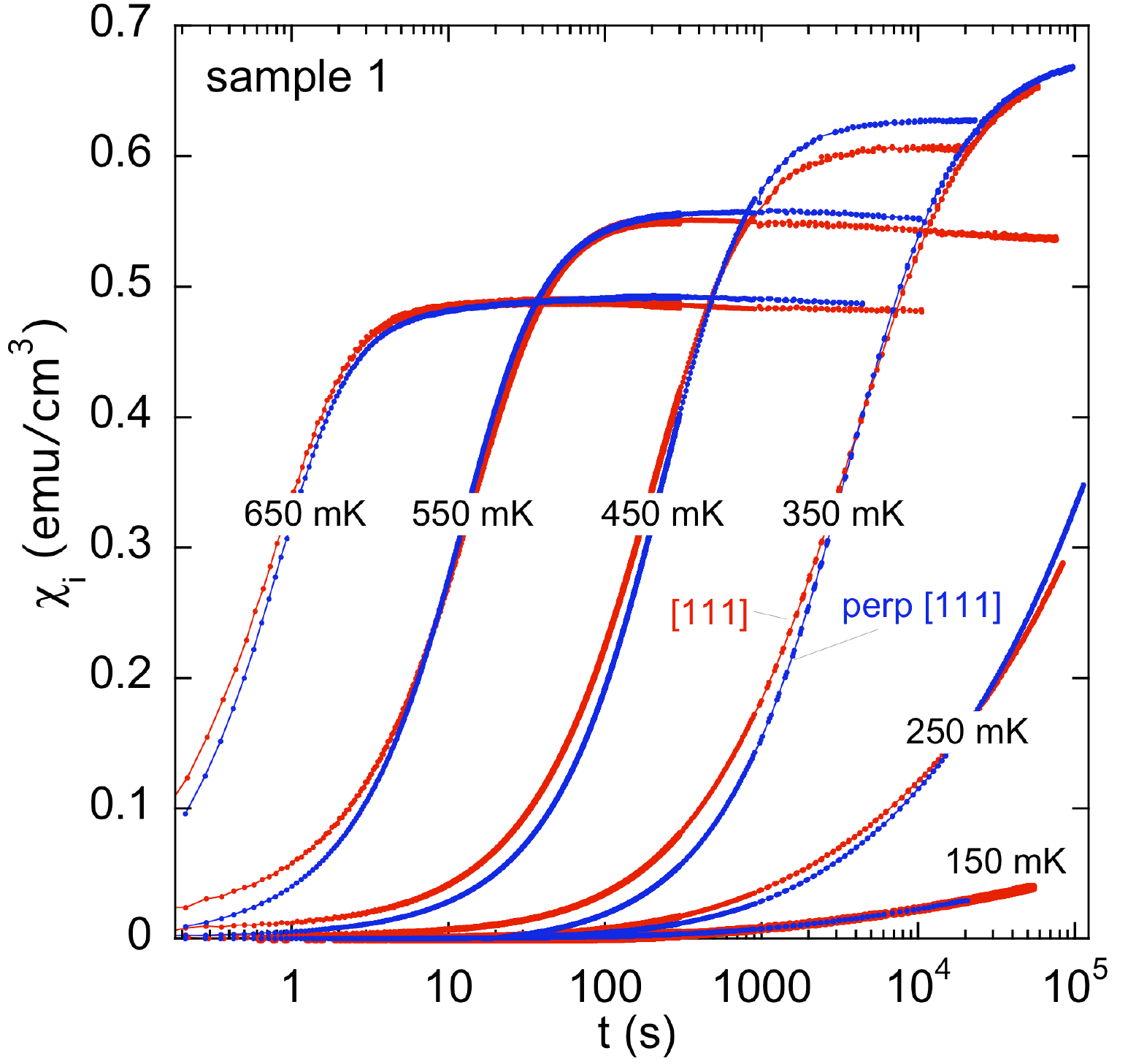}
\caption{({\bf left}) Relaxation of the magnetization measured at 350 mK for sample 1 with the field applied perpendicular to the [111] direction. The sample was first ZFC, with cooling rates from 10 mK/s to 0.09 mK/s as shown in the figure.  Then a field of 50 Oe was applied and the resulting relaxation of the magnetization was recorded. Also shown is the relaxation using the avalanche quench protocol for the same conditions. ({\bf right}) Relaxation at various temperatures using the avalanche quench protocol as described in the figure 1 for the field along the [111] direction (red) and for the field perpendicular to the [111] direction (blue), for the same sample. Plotted is the dc susceptibility $\chi_i= M/H_i$ where $H_i$ is the demagnetization corrected internal field, necessary in order to take into account the different demagnetization coefficients for the two directions. As can be seen, the relaxation along the two directions is very similar, with a slight preference for the [111] direction. } 
\label{fig_S4}
\end{figure}

Regardless of the sample, the effects of cooling rate were very strong. The faster the cooling, the faster the resulting relaxation, and the effect becomes more marked for lower temperatures.  In this respect, figure \ref{fig_S3} shows that although sample 3, $H$// [001] was slower, it behaved like samples 1 and 2 with respect to cooling rates. 
A rather weak dependence of the direction of the applied field was observed on the relaxation for samples 1 and 3. This is shown in figure \ref{fig_S4} for the demagnetization corrected dc susceptibility for sample 1 measured along the [111] direction, and perpendicular to [111].  Nevertheless, for both samples the [111] direction appeared slightly faster. Also shown in the figure is the relaxation using the avalanche quench protocol, which always resulted in the fastest relaxation.

\paragraph*{1.4. Comments on the Avalanche Quench protocol:}

In all cases the AQ protocol resulted in significantly faster relaxation, and thus presumably a higher concentration of monopoles at the start of the relaxation. The objective of this technique is to make a rapid thermal quench of the sample with the fastest $dT/dt$ possible, in order to freeze in as many monopoles as possible. To do this, we use a trick: we use the magnetic energy $\Delta M \cdot H$ that is liberated when the magnetization $M$ flips abruptly in a field $H$, akin to the magnetic avalanche. This energy will at first just heat the sample more or less adiabatically, and the temperature inside the sample rises very fast, presumably above 900 mK (as when cooled after the AQ in zero field the magnetization approaches zero -see figure 1- indicating the sample has been warmed well above the known freezing temperature\cite{SnyderSchiffer}). 
The heat will then rapidly leak out of the sample, and eventually be absorbed by the sample holder and transferred to the $^3$He in the mixing chamber. At low temperature, the specific heat of $^3$He is three orders of magnitude greater than everything else in the system, so the resulting temperature increase is small (even for our miniature dilution system), of the order $\Delta T=200$ mK depending on the sample size etc. This is the key result, the mixing chamber remains cold, so that after the avalanche, the sample will cool back down extremely quickly. 

The cool down is governed by the thermal conduction inside the sample and the thermal contact with the mixing chamber. The thermometer is attached to the mixing chamber which is 25 cm from the sample position. In our case the thermal contact is very good because the sample is sandwiched between two copper plates that are coated with a small film of apiezon grease, and are bolted to the mixing chamber. Thus depending on the sample, the conduction occurs over a very large surface area. From extrapolations of our controlled cooling rate data from the conventional relaxation measurements, we  estimate the cooling in the avalanche quench to be 0.06 K/s or faster. 

We tested various avalanching  protocols before eventually settling on the method outlined in figure 1. For this protocol, we varied the avalanche field up to 4000 Oe, and found no significant difference with field. On the other hand when $H=1500$ Oe, the avalanches were sometimes erratic, and for smaller fields, avalanches  did not occur at all at very low temperature. 


It may seem strange that reducing the field from 2000 Oe to zero results in heating of the sample, which is at odds with the well know procedure for cooling a sample using adiabatic demagnetization. There are of course many differences. First, spin ice is a system of  strongly correlated frustrated spins. When the field is changed very rapidly, the spins do not have time to flip. In a sense we have Òpulled the rug out from underneath the spinsÓ, leaving them ÒstandingÓ. The spins will then ultimately see the oppositely pointing internal field which for our samples was considerable, 500 to 800 Oe depending on the sample, and when they flip, it is this field that will give rise to the energy. Indeed, Orend\'{a}\v{c} and co-workers \cite{martin} working with a powder sample of Dy$_2$Ti$_2$O$_7$ starting from 850 mK could cool below 300 mK by adiabatic demagnetization, using  a field sweep rate of 0.02 Oe/s. This is approximately $2.7 \times 10^5$ slower than the sweep rate used in this work.

\paragraph*{1.5. Magnetization dynamics in the Dipolar Spin Ice model:}  

Whilst the near neighbour spin ice model captures the essential and qualitative features of the low temperature magnetization dynamics, it is worth mentioning that the more sophisticated dipolar spin ice (DSI) model captures more quantitatively the observed properties. 

\begin{figure}
\center{
\includegraphics[keepaspectratio=true, width=15cm]{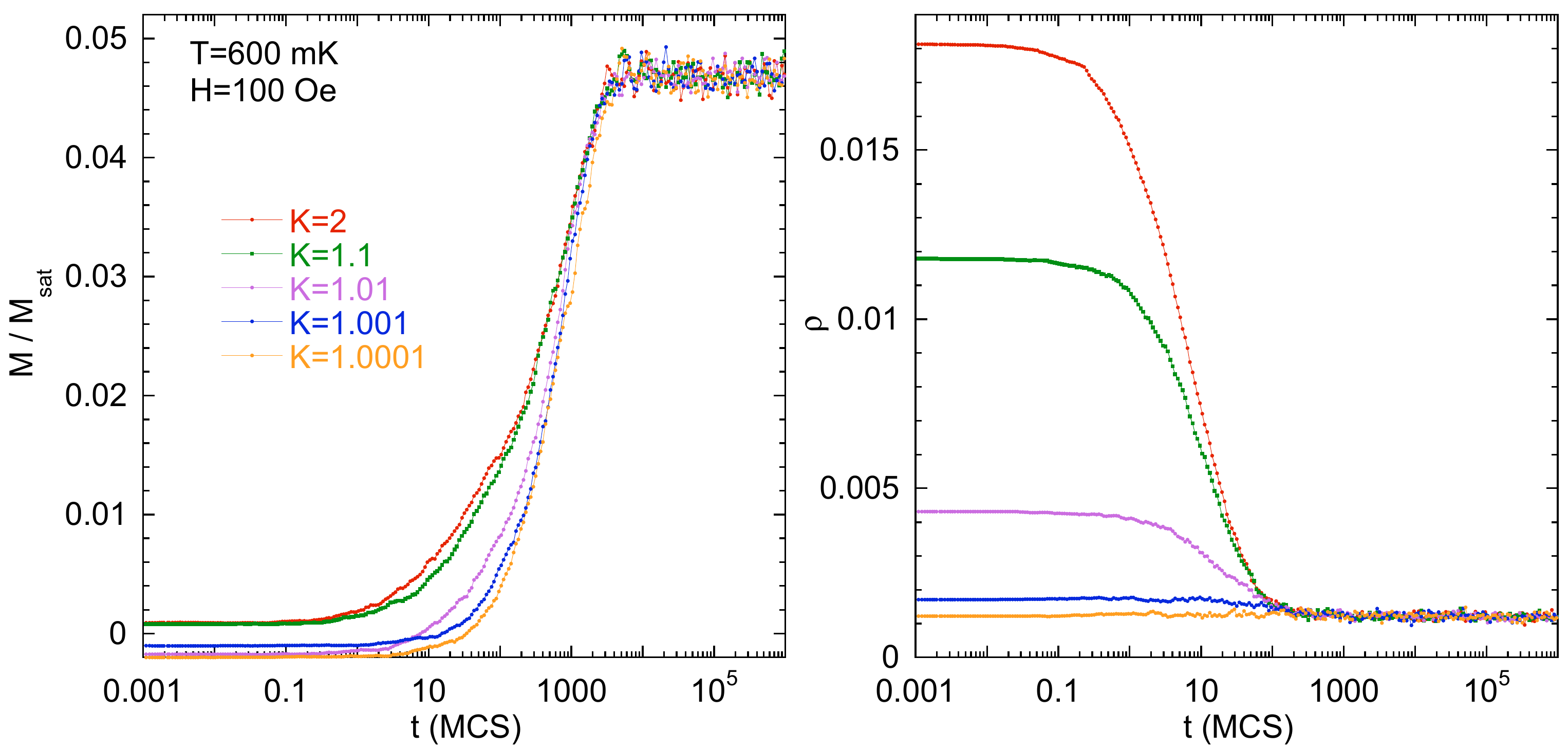}}
\caption{Relaxation of the magnetization ({\bf left}) and density of monopoles $\rho$ ({\bf right}) vs time $t$ at 600 mK in the dipolar spin ice model (see text).} 
\label{fig_S5}
\end{figure}

\begin{figure}
\center{
\includegraphics[keepaspectratio=true, width=15cm]{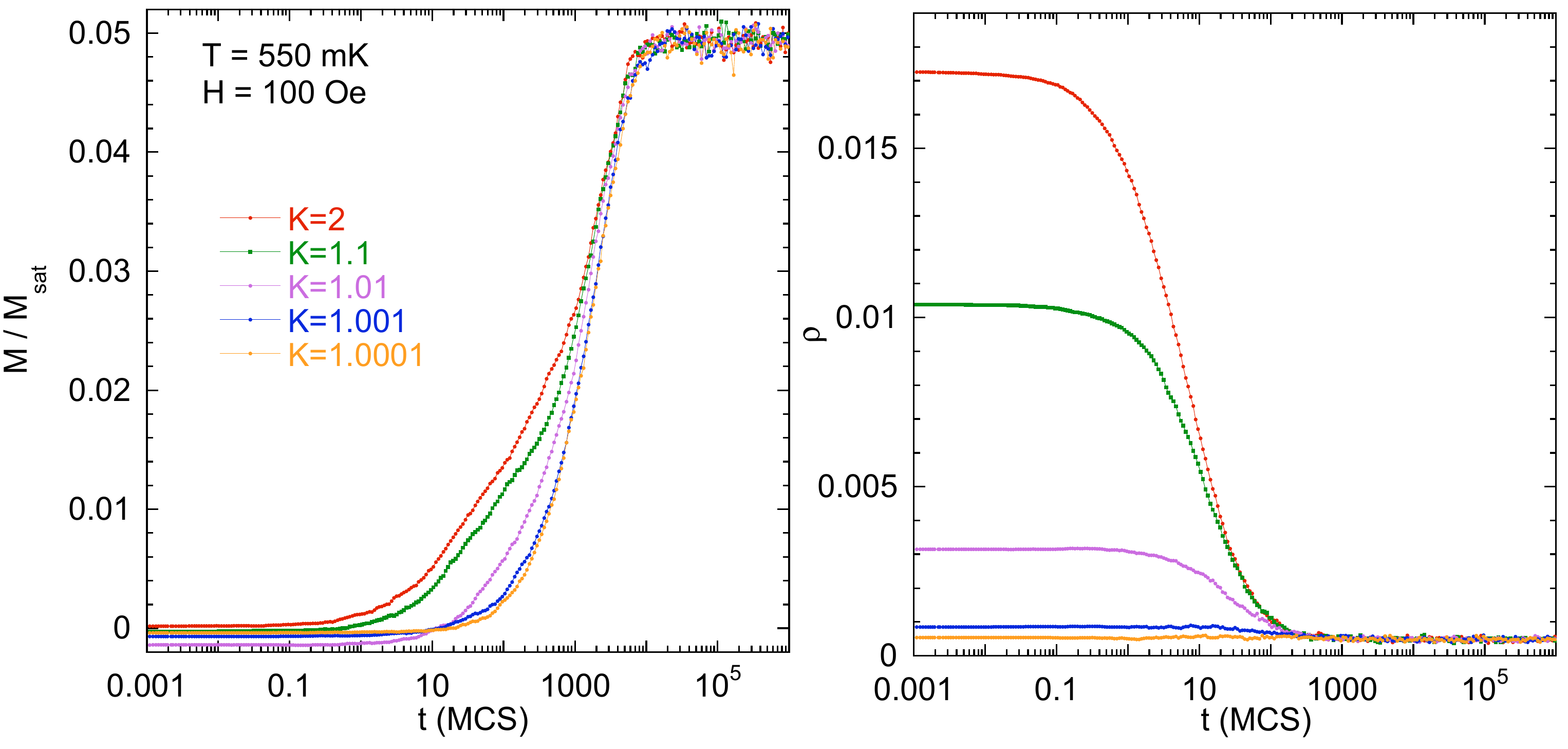}}
\caption{Relaxation of the magnetization ({\bf left}) and density of monopoles $\rho$ ({\bf right}) vs time $t$ at 550 mK in the dipolar spin ice model (see text).} 
\label{fig_S6}
\end{figure}

Figure \ref{fig_S5} and \ref{fig_S6} show $M$ (left) and monopole density $\rho$ (right) vs time $t$ in a semilogarithmic scale, from a kinetic Monte Carlo simulation of a dipolar spin ice model of Dy$_2$Ti$_2$O$_7$. Simulations are performed on $L \times L\times L \times 16$ samples with $L=$8 (ie 8192 sites), periodic boundary conditions, and long range interactions are taken into account through the Ewald summation technique \cite{Melko} while spin dynamics are described with a single spin flip algorithm. 

As for the near neighbour model, the model is thermally quenched by reducing the temperature in equal logarithmic steps, from 1 K to 600 (see figure \ref{fig_S5}) or 550 mK (see figure \ref{fig_S6}), in zero applied magnetic field. The logarithmic decrement, from right to left is $K=1.0001$, 1.001, 1.01, 1.1 and 2. A field of 100 Oe is then applied at $t = 0$ along the [111] direction and $M(t)$ is monitored until it reaches its asymptotic, cooling rate independent, value. The microscopic parameters for the simulation are those of Dy$_2$Ti$_2$O$_7$ (see Ref \citenum{Jaubert} and associated references). Each $M(t)$ curve corresponds to an average of 100 independent cooling scenarios. Time is given in units of Monte Carlo Steps (MCS), 1 MCS being associated to a stochastic sampling of the whole sample in average.

Because long range interactions shift the dynamics to very long time, we restrict ourselves to relatively high temperatures, i.e around 500 mK, in order to computationally work in an accessible time window. This point being made, it is worth emphasizing that our time scale, the MCS, is an intrinsic time scale, i.e a time related to the single spin flip (SSF) dynamics of our statistical spin model. The experimental one is impeded by another temperature dependent factor, which would be, for instance, an exponential one, $\tau_0  \exp(E/k_BT)$, provided one assumes thermal activated process for the SSF. In other words, the dynamical slowing down we are dealing with for the DSI model is related to the SSF dynamics of the spin model in its free energy landscape, not another microscopic process.

\paragraph*{1.6. Magnetization dynamics in the Near Neighbour Spin Ice model:} 

\begin{figure}[h!]
\center{
\includegraphics[keepaspectratio=true, width=8cm]{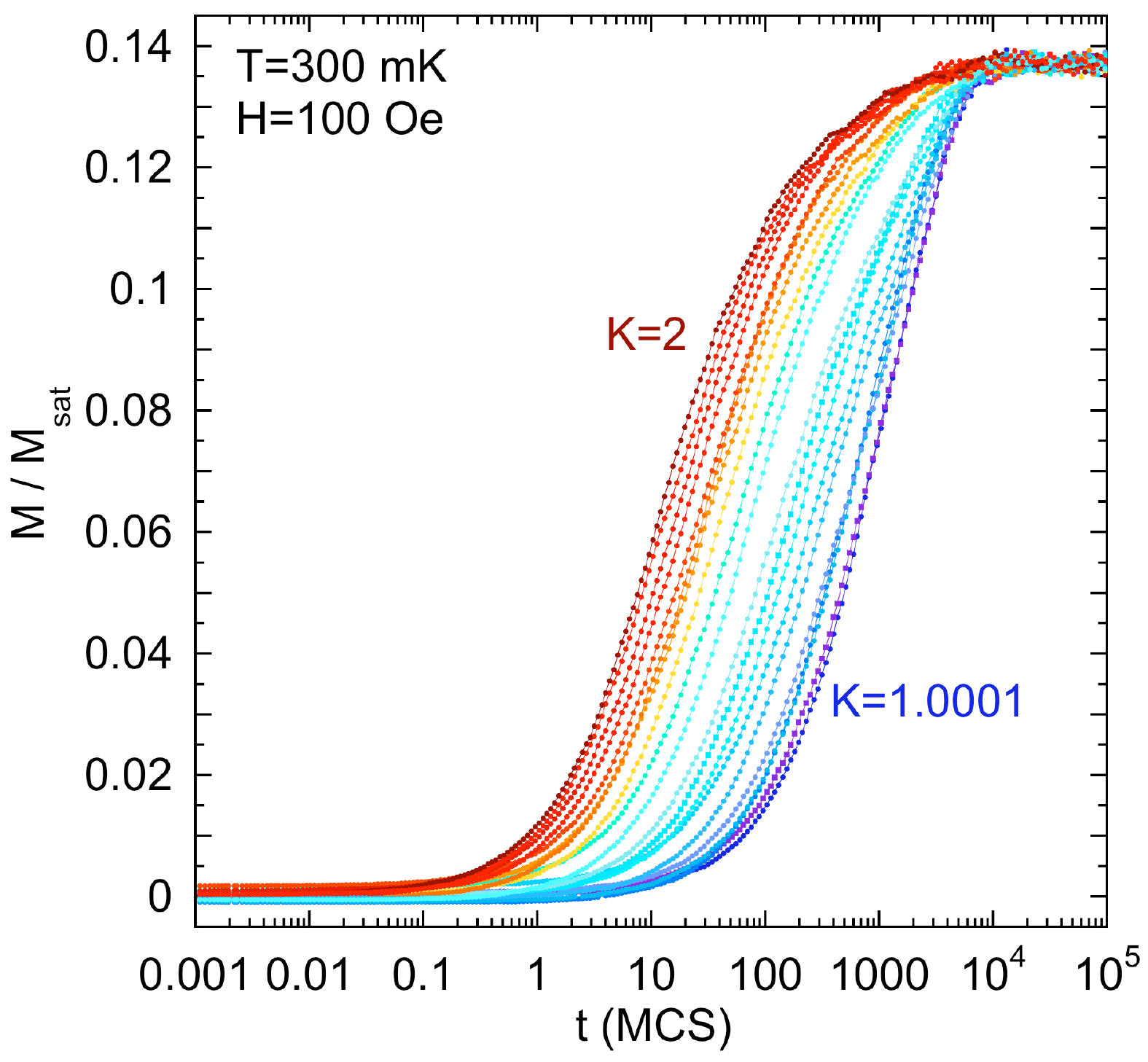}
\caption{Relaxation of the magnetization vs time $t$ at 300 mK in the nearest neighbour spin ice model (see text).} 
\label{fig_S7}}
\end{figure}

\begin{figure}[h!]
\center{
\includegraphics[keepaspectratio=true, width=7.5cm]{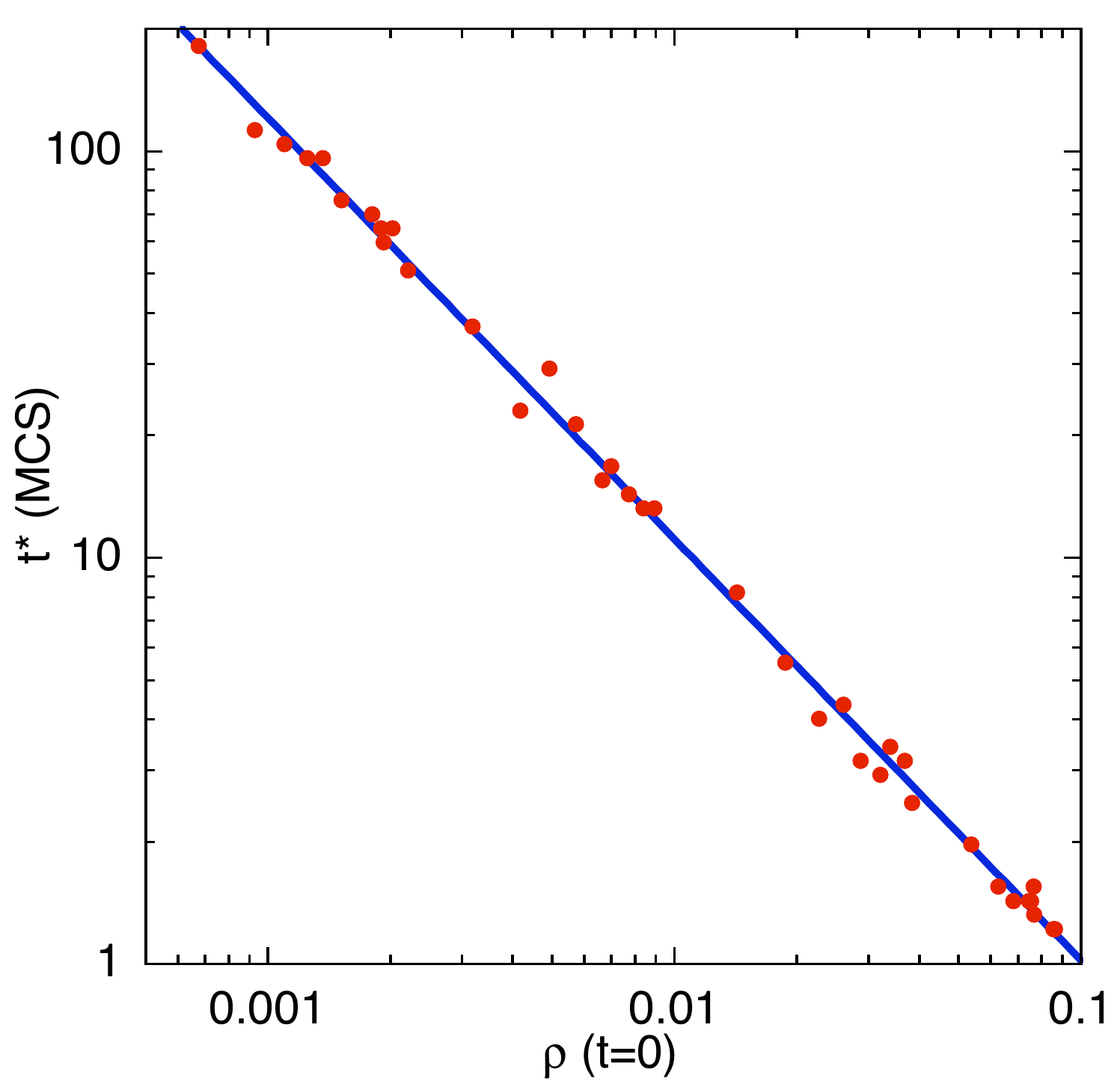}
\caption{$t^*$ vs $\rho(t=0)$ at 300 mK in the nearest neighbour spin ice model (see text).} 
\label{fig_S8}}
\end{figure}

On the other hand, it is possible to investigate a larger set of cooling rates in the near neighbour case. Fig \ref{fig_S7} shows a large set (38 cooling rates distributed logarithmically from K=0.0001 to K=2) of kinetic monte carlo simulations of the near neighbour model at 300 mK as described in the manuscript. 

From those simulations, the time at which $M(t)$ reaches one tenth of the asymptotic value $M(\infty)$ is related to the $t=0$ value of the monopole density, i.e the monopole density just before applying the external magnetic field. Fig \ref{fig_S8} (or inset of fig 3) is a log log plot of this dependence, and can be fitted by $t^* = A/(\rho(t=0))^b$, with ($A=0.059$ and $b=1.03$).


\end{document}